\documentclass[amsmath,aps,showpacs,a4paper,10pt]{revtex4}

 \usepackage{epsf}
 \usepackage{graphicx}    

 \usepackage{enumerate}   

\begin{document}

 \newcommand{\be}[1]{\begin{equation}\label{#1}}
 \newcommand{\ee}{\end{equation}}
 \newcommand{\bea}{\begin{eqnarray}}
 \newcommand{\eea}{\end{eqnarray}}
 \def\disp{\displaystyle}

 \def\gsim{ \lower .75ex \hbox{$\sim$} \llap{\raise .27ex \hbox{$>$}} }
 \def\lsim{ \lower .75ex \hbox{$\sim$} \llap{\raise .27ex \hbox{$<$}} }

 \begin{titlepage}

 \begin{flushright}
 arXiv:1804.03087
 \end{flushright}

 \title{\Large \bf Testing the Cosmic Anisotropy with Supernovae Data:
 Hemisphere~Comparison and Dipole Fitting}

 \author{Hua-Kai~Deng\,}
 \email[\,email address:\ ]{dhklook@163.com}
 \affiliation{School of Physics,
 Beijing Institute of Technology, Beijing 100081, China}

 \author{Hao~Wei\,}
 \email[\,Corresponding author;\ email address:\ ]{haowei@bit.edu.cn}
 \affiliation{School of Physics,
 Beijing Institute of Technology, Beijing 100081, China}

 \begin{abstract}\vspace{1cm}
 \centerline{\bf ABSTRACT}\vspace{2mm}
 The cosmological principle is one of the cornerstones in
 modern cosmology. It assumes that the universe is homogeneous
 and isotropic on cosmic scales. Both the homogeneity and the
 isotropy of the universe should be tested carefully. In the
 present work, we are interested in probing the possible
 preferred direction in the distribution of type Ia supernovae
 (SNIa). To our best knowledge, two main methods have been used
 in almost all of the relevant works in the literature, namely
 the hemisphere comparison (HC) method and the dipole fitting
 (DF) method. However, the results from these two methods are
 not always approximately coincident with each other. In this
 work, we test the cosmic anisotropy by using these two methods
 with the Joint Light-Curve Analysis (JLA) and simulated SNIa
 datasets. In many cases, both methods work well, and their
 results are consistent with each other. However, in the cases
 with two (or even more) preferred directions, the DF method
 fails while the HC method still works well. This might
 shed new light on our understanding of these two methods.
 \end{abstract}

 \pacs{98.80.-k, 98.80.Es, 95.36.+x}

 \maketitle

 \end{titlepage}

 \renewcommand{\baselinestretch}{1.0}


\section{Introduction}\label{sec1}

As is well known, the cosmological principle
 is one of the cornerstones in modern cosmology~\cite{Weinberg:cos,
 Kolb90}. It assumes that the universe is homogeneous and isotropic
 on cosmic scales. In fact, the cosmological principle has been
 observed to be approximately valid across a very large part
 of the universe (e.g.~\cite{Hogg:2004vw,Hajian:2006ud}).
 However, it is not born to be true, and this assumption should
 be strictly tested by using the cosmological observations. As
 its two main parts, both the homogeneity and the isotropy of
 the universe should be probed carefully.

In fact, the cosmological principle has not yet been well proven on
 cosmic scales $\gsim\,1\,$Gpc~\cite{Caldwell:2007yu}. On the
 other hand, the local universe is obviously inhomogeneous and
 anisotropic on small scales. In particular, the nearby sample
 has been examined for evidence of a local
 ``Hubble Bubble''~\cite{Zehavi:1998gz}. If the cosmological
 principle can be relaxed, it is possible to explain the
 apparent cosmic acceleration discovered in 1998~\cite{Riess:1998cb,
 Perlmutter:1998np}, without invoking dark energy~\cite{DE} or
 modified gravity~\cite{MG}. For instance, giving up the cosmic
 homogeneity, it is reasonable to imagine that we are living
 in a locally underdense void. One of such models is the well-known
 Lema\^{\i}tre-Tolman-Bondi~(LTB) void model~\cite{LTB,
 Goode:1982pg,Alnes:2005rw,GarciaBellido:2008nz,Enqvist:2006cg,
 Celerier:2012xr,Ishak:2013vha,Clifton:2008hv,Zhang:2012qr,
 Yan:2014eca}. In this model, the universe is spherically symmetric
 and radially inhomogeneous, and we are living in a locally
 underdense void centered nearby our location. The Hubble
 diagram inferred from lines-of-sight originating at the center
 of the void might be misinterpreted to indicate cosmic
 acceleration~\cite{Alnes:2005rw,GarciaBellido:2008nz,
 Enqvist:2006cg,Celerier:2012xr,Ishak:2013vha,Clifton:2008hv,
 Zhang:2012qr,Yan:2014eca}. In fact, the LTB-like
 models violating the cosmological principle have
 been extensively considered in the literature nowadays.

In the literature, the cosmic homogeneity has been tested by
 using e.g. type~Ia supernovae~(SNIa)~\cite{Celerier:1999hp,
 Clifton:2008hv,Celerier:2012xr}, cosmic microwave background
 (CMB)~\cite{Caldwell:2007yu,Moffat:2005yx,Alnes:2006pf,
 Clifton:2009kx,Clarkson:2010ej,Moss:2010jx}, time drift of
 cosmological redshifts~\cite{Uzan:2008qp,Quartin:2009xr},
 baryon acoustic oscillations (BAO)~\cite{Bolejko:2008cm,
 February:2012fp,Zibin:2008vk}, integrated Sachs-Wolfe
 effect~\cite{Tomita:2009wz},
 galaxy surveys~\cite{Labini:2010qx}, kinetic
 Sunyaev Zel'dovich effect~\cite{Zhang:2010fa,
 Valkenburg:2012td,Bull:2011wi,Moss:2011ze,Marra:2011ct}, ages
 of old high-redshift objects~\cite{Yan:2014eca}, observational
 $H(z)$ data~\cite{Zhang:2012qr}, and growth of large-scale
 structure~\cite{Celerier:2012xr}. However, the debate on
 the inhomogeneous universe has not been settled by now.

In contrast to the LTB-like models giving up the cosmic homogeneity,
 there is another kind of models violating the cosmological
 principle, in which the universe is not isotropic. For example, the
 well-known G\"odel solution~\cite{Godel:1949ga} of the
 Einstein field equations describes a homogeneous rotating universe.
 Although the G\"odel universe has some exotic features (see
 e.g.~\cite{Li:2016nnn}), it is indeed an interesting idea that
 our universe is rotating around an axis. In fact, this idea can be
 completely independent of the G\"odel universe. In addition, there
 are other kinds of anisotropic models in the literature. For
 instance, most of the well-known Bianchi type I $\sim$ IX
 universes~\cite{Bianchi,Mishra:2015jja} are anisotropic in general.

In fact, some hints of the cosmic anisotropy have been claimed
 in the literature. For example, it is found that there exists
 a preferred direction in the CMB temperature map (known as
 the ``Axis~of~Evil'' in the literature)~\cite{Axisofevil,
 Zhao:2016fas,Hansen:2004vq}, the distribution of
 SNIa~\cite{Schwarz:2007wf,Antoniou:2010gw,Mariano:2012wx,Cai:2011xs,
 Zhao:2013yaa,Yang:2013gea,Chang:2014nca,Lin:2015rza,Chang:2017bbi,
 Javanmardi:2015sfa,Bengaly:2015dza}, gamma-ray bursts
 (GRBs)~\cite{Meszaros:2009ux,Wang:2014vqa,Chang:2014jza},
 rotationally supported galaxies~\cite{Zhou:2017lwy,Chang:2018vxs},
 quasars and radio galaxies~\cite{Singal:2013aga,Bengaly:2017slg}, and
 the quasar optical polarization
 data~\cite{Hutsemekers,Pelgrims:2016mhx}. In addition, using
 the absorption systems in the spectra of distant quasars, it
 is claimed that the fine structure ``constant'' $\alpha$ is
 not only time-varying~\cite{Webb98,Webb00} (see
 also e.g.~\cite{{Uzan10,Barrow09,HWalpha}}), but also
 spatially varying~\cite{King:2012id,Webb:2010hc}. Precisely
 speaking, there is also a preferred direction in the data of
 $\Delta\alpha/\alpha$. It is found in~\cite{Mariano:2012wx}
 that the preferred direction in $\Delta\alpha/\alpha$ might be
 correlated with the one in the distribution of SNIa. Up
 to date, the hints of the cosmic anisotropy are still accumulating.

In the present work, we are interested in probing the possible
 preferred direction in the distribution of
 SNIa~\cite{Schwarz:2007wf,Antoniou:2010gw,Mariano:2012wx,Cai:2011xs,
 Zhao:2013yaa,Yang:2013gea,Chang:2014nca,Lin:2015rza,Chang:2017bbi}.
 To our best knowledge, two main methods have been used in almost all
 of the relevant works in the literature (e.g.~\cite{Schwarz:2007wf,
 Antoniou:2010gw,Mariano:2012wx,Cai:2011xs,Zhao:2013yaa,Yang:2013gea,
 Chang:2014nca,Lin:2015rza,Chang:2017bbi}), namely the
 hemisphere comparison (HC) method proposed in~\cite{Schwarz:2007wf}
 and then improved by~\cite{Antoniou:2010gw} (see also
 e.g.~\cite{Cai:2011xs,Yang:2013gea,Chang:2014nca}), and the dipole
 fitting (DF) method proposed in~\cite{Mariano:2012wx} (see
 also e.g.~\cite{Chang:2014nca,Lin:2015rza,Chang:2017bbi,
 Zhou:2017lwy,Chang:2018vxs,Yang:2013gea,Wang:2014vqa}). In the
 HC method, the data points are randomly divided into many pairs of
 hemispheres according to their positions in the sky, and then
 these pairs of hemispheres are compared until the preferred
 direction with a maximum anisotropy level is found. In the DF
 method, the data points are directly fitted to a dipole (or
 dipole plus monopole in some cases). We refer to the next
 sections for the details of these two methods.

It is natural to expect that the preferred directions found by
 these two methods are approximately coincident with each other. Of
 course, in many cases the answer is ``yes''. However, it is
 not always ``yes'' unfortunately. For example, the preferred
 direction in the Union2 SNIa dataset found by the DF method is
 approximately opposite to the one found by the
 HC method~\cite{Chang:2014nca}. On the other hand, a preferred
 direction in the Union2.1 SNIa dataset was found by the DF
 method, but there is a null signal for the HC
 method~\cite{Yang:2013gea}. In addition, the DF method failed
 to find the preferred direction in the JLA SNIa
 dataset~\cite{Lin:2015rza,Chang:2017bbi}. To our best
 knowledge, the HC method has not been used to find the
 preferred direction in the JLA SNIa dataset up to now, and
 hence we do this in the present work. In contrast to the
 failure of the DF method~\cite{Lin:2015rza,Chang:2017bbi},
 the HC method works well in the JLA SNIa dataset (see below).
 Therefore, it is of interest to compare these two methods
 carefully, and we will do this by using several simulated SNIa
 datasets. In fact, this might shed new light
 on our understanding of these two methods.

The rest of this paper is organized as follows. In Sec.~\ref{sec2},
 we briefly review the key points of the HC method and the DF
 method, and then we use them to find the possible preferred
 direction in the JLA SNIa dataset. In Sec.~\ref{sec3}, we
 compare these two methods by using several simulated SNIa datasets.
 In Sec.~\ref{sec4}, some brief concluding remarks are given.


\section{The preferred direction in the JLA SNIa dataset}\label{sec2}

As mentioned above, to our best knowledge, the HC method has
 not been used to find the preferred direction in the JLA
 dataset consisting of 740 SNIa~\cite{Betoule:2014frx} up to
 now. We will do this here. At first, we briefly review the
 key points of the HC method following~\cite{Antoniou:2010gw}.
 Its goal is to identify the direction of the axis of maximal
 asymmetry for the corresponding dataset. Usually, the physical
 quantity to be compared is the accelerating expansion rate,
 namely the deceleration parameter $q_0$~\cite{Weinberg:cos,Kolb90}
 (note that $q_0<0$ means that the universe is accelerating).
 As is well known, in the spatially flat $\Lambda$CDM model,
 the deceleration parameter $q_0$ is related to the fractional
 density of the pressureless matter $\Omega_{m0}$ according to
 $q_0=-1+3\Omega_{m0}/2$. So, it is convenient to use $\Omega_{m0}$
 instead~\cite{Antoniou:2010gw}, as we consider the spatially
 flat $\Lambda$CDM model throughout this work.
 Following~\cite{Antoniou:2010gw}, the main steps to implement
 the HC method are ({\it i}) Generate a random direction
 $\hat{r}_{\rm rnd}$ indicated by $(l,\,b)$ with a uniform
 probability distribution, where $l\in [\,0^\circ,\,360^\circ)$ and
 $b\in [-90^\circ,\,+90^\circ\,]$ are the longitude and the
 latitude in the galactic coordinate system, respectively.
 ({\it ii}) Split the dataset under consideration into two
 subsets according to the sign of the inner
 product $\hat{r}_{\rm rnd}\cdot\hat{r}_{\rm dat}$, where
 $\hat{r}_{\rm dat}$ is a unit vector describing the direction
 of each SNIa in the dataset. Thus, one subset corresponds to
 the hemisphere in the direction of the random vector (defined
 as ``up''), while the other subset corresponds to the opposite
 hemisphere (defined as ``down''). Noting that the position of
 each SNIa in the dataset is usually given by right ascension
 (ra) and declination (dec) in degree (equatorial coordinate
 system, J2000), one should convert $\hat{r}_{\rm rnd}$ and
 $\hat{r}_{\rm dat}$ to Cartesian coordinates in this step.
 ({\it iii}) Find the best-fit values on $\Omega_{m0}$ in each
 hemisphere ($\Omega_{m0,u}$ and $\Omega_{m0,d}$), and then
 obtain the so-called anisotropy level (AL) quantified through
 the normalized difference~\cite{Antoniou:2010gw},
 \be{eq1}
 {\rm AL}\equiv\frac{\Delta\Omega_{m0}}{\bar{\Omega}_{m0}}=
 2\cdot\frac{\Omega_{m0,u}-\Omega_{m0,d}}{\Omega_{m0,u}
 +\Omega_{m0,d}}\,.
 \ee
 ({\it iv}) Repeat for $N$ random directions $\hat{r}_{\rm rnd}$ and
 find the maximum AL, as well as the corresponding direction of
 maximum anisotropy. ({\it v}) Obtain the
 $1\sigma$ error $\sigma_{\rm AL}$ associated with the
 maximum AL~\cite{Antoniou:2010gw},
 \be{eq2}
 \sigma_{\rm AL}=\frac{\sqrt{\sigma_{\Omega_{m0,u}^{\rm max}}^2
 +\sigma_{\Omega_{m0,d}^{\rm max}}^2}}{\Omega_{m0,u}^{\rm max}
 +\Omega_{m0,d}^{\rm max}}\,.
 \ee
 Note in~\cite{Antoniou:2010gw} that $\sigma_{\rm AL}$ is the
 error due to the uncertainties of the SNIa distance moduli
 propagated to the best-fit $\Omega_{m0}$ on each hemisphere
 and thus to AL. One can identify all the test axes corresponding to
 an AL within $1\sigma$ from the maximum AL,
 namely ${\rm AL = AL}_{\rm max}\pm\sigma_{\rm AL}$. These axes
 cover an angular region corresponding to the $1\sigma$ range
 of the maximum anisotropy direction. We refer
 to~\cite{Antoniou:2010gw} for more details of the HC method.

In many of the relevant works following~\cite{Antoniou:2010gw},
 Mathematica was commonly used, and the number of random directions
 in step ({\it iv}) are taken to be approximately equal to the
 number of data points as suggested by~\cite{Antoniou:2010gw}.
 In this work, we use Matlab instead, and the number of random
 directions in step ({\it iv}) can be $N\sim{\cal O}(10^4)$ or
 even more.

\begin{center}
 \begin{figure}[tb]
 \centering
 \vspace{-3mm}  
 \includegraphics[width=0.95\textwidth]{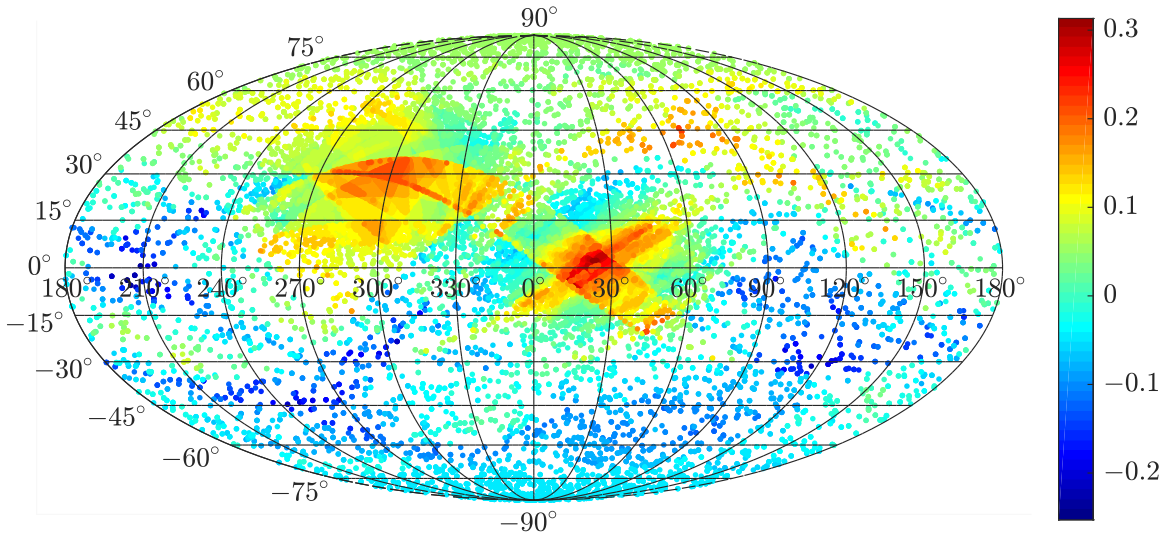}
 \caption{\label{fig1} The pseudo-color map of AL$(l,\,b)$ obtained
 by using the HC method to the JLA SNIa dataset. The two preferred
 directions $(23.49^\circ,\,2.25^\circ)$ and
 $(299.47^\circ,\,28.39^\circ)$ are within the red regions. See the
 text for details.}
 \end{figure}
 \end{center}

Here, we implement the HC method to the JLA dataset consisting
 of 740 SNIa~\cite{Betoule:2014frx}. We first repeat 10000
 random directions $(l,\,b)$ across the whole sky, and find that the
 directions with the largest ALs concentrate around two
 directions, namely $(300.6575^\circ,\,28.1678^\circ)$ and
 $(23.4274^\circ,\,1.7021^\circ)$. Then, we densely repeat
 $5000 \sim 20000$ random directions from the Gaussian
 distributions with the means in these two preliminary
 directions, respectively. Finally, we find that the $1\sigma$
 angular region with the maximum AL is in the direction
 \be{eq3}
 (l,\,b)\,_{\rm HC,\,max}^{\rm JLA}
 =({23.4893^\circ}\,^{+21.6274^\circ}_{-12.8318^\circ}\,,\,
 {2.2524^\circ}\,^{+3.7961^\circ}_{-22.6837^\circ})\,,
 \ee
 and the corresponding maximum AL (with $1\sigma$
 uncertainty) is
 \be{eq4}
 {\rm AL_{max}^{JLA}}=0.3132\pm 0.1003\,.
 \ee
 In addition, we also find a sub-maximum AL in the direction
 (with $1\sigma$ uncertainty)
 \be{eq5}
 (l,\,b)\,_{\rm HC,\,sub}^{\rm JLA}
 =({299.4711^\circ}\,^{+46.1314^\circ}_{-23.3855^\circ}\,,\,
 {28.3912^\circ}\,^{+6.5202^\circ}_{-17.0096^\circ})\,,
 \ee
 and the corresponding sub-maximum AL (with $1\sigma$ uncertainty) is
 \be{eq6}
 {\rm AL_{sub}^{JLA}}=0.2873\pm 0.1110\,.
 \ee
 In fact, it is not so rare to find two preferred directions
 (see e.g.~\cite{Zhou:2017lwy}). Note that the second preferred
 direction given in Eq.~(\ref{eq5}) is consistent with the one
 $(l,\,b)=(309^\circ,\,18^\circ)$ found in~\cite{Antoniou:2010gw} for
 the Union2 SNIa dataset within the $1\sigma$ region. We present the
 pseudo-color map of AL$(l,\,b)$ in Fig.~\ref{fig1}. It is clear to
 see the two preferred directions within the red regions.


\vspace{-12mm}  


 \begin{center}
 \begin{figure}[tb]
 \centering
 \vspace{-4mm}  
 \includegraphics[width=1.0\textwidth]{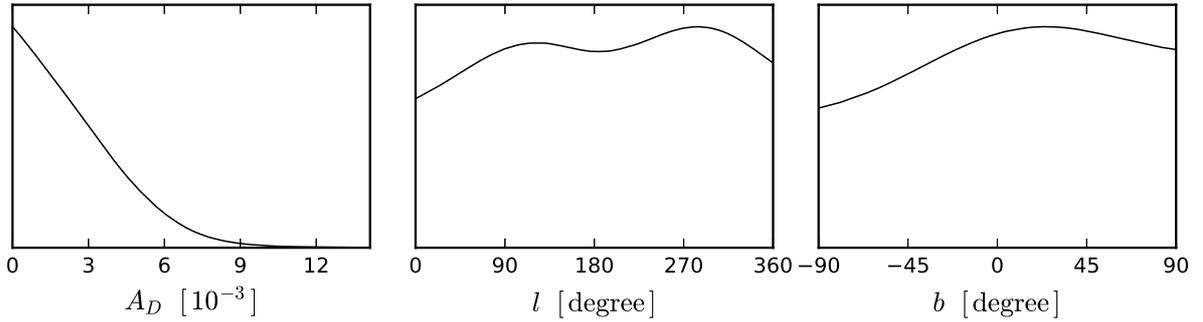}
 \caption{\label{fig2} The marginalized probability distributions of
 the dipole magnitude $A_D$ and the dipole direction $(l,\,b)$,
 obtained by using the DF method to the JLA SNIa dataset with
 the priors $\Omega_{m0}=0.295$ and $A_D\geq 0$. Note that
 $A_D$ is given in units of $10^{-3}$. See the text for
 details.}
 \end{figure}
 \end{center}


\vspace{-10mm}  

Next, let us turn to the DF method. It has already been known
 that the DF method failed to find the preferred direction in
 the JLA SNIa dataset~\cite{Lin:2015rza,Chang:2017bbi}. But
 here we would like to generalize the main results. At first,
 we briefly review the key points of the DF method following
 e.g.~\cite{Mariano:2012wx,Chang:2014nca,Lin:2015rza,Chang:2017bbi,
 Zhou:2017lwy,Chang:2018vxs,Yang:2013gea,Wang:2014vqa}. If the
 observational quantity under consideration is denoted by
 $\xi$, the corresponding $\chi^2$ is given by
 $\chi^2=(\vec{\xi}_{obs}-\vec{\xi}_{th})^T \,{\bf C}^{-1}
 (\vec{\xi}_{obs}-\vec{\xi}_{th})$, where $\bf C$ is the
 covariance matrix of $\vec{\xi}$. When $\bf C$ is a diagonal
 matrix, it reduces to $\chi^2=\sum (\xi_{obs,i}-
 \xi_{th,i})^2/\sigma_{\xi,i}^2\,$. If $\xi$ is anisotropic,
 one can consider a dipole plus monopole correction, namely
 $\xi_{th}=\bar{\xi}_{th}\left[\,1+B
 +A_D\left(\hat{n}\cdot\hat{p}\right)\,\right]$, where $B$
 and $A_D$ are the monopole term and the dipole magnitude,
 respectively; $\hat{n}$ is the dipole direction; $\hat{p}$ is
 the unit 3-vector pointing toward the data point; $\bar{\xi}_{th}$
 is the value predicted by the isotropic theoretical model.
 Usually, the monopole term $B$ is negligible, and one can
 only consider the dipole modulation, namely
 \be{eq7}
 \xi_{th}=\bar{\xi}_{th}\left[\,1
 +A_D\left(\hat{n}\cdot\hat{p}\right)\,\right].
 \ee
 In terms of the galactic coordinates $(l,\,b)$, the dipole
 direction is given by
 \be{eq8}
 \hat{n}=\cos(b)\cos(l)\,\hat{\bf i}+\cos(b)
 \sin(l)\,\hat{\bf j}+\sin(b)\,\hat{\bf k}\,,
 \ee
 where $\hat{\bf i}$, $\hat{\bf j}$, $\hat{\bf k}$ are the unit
 vectors along the axes of Cartesian coordinates system. The
 position of the $i$-th data point with the
 galactic coordinates $(l_i,\,b_i)$ is given by
 \be{eq9}
 \hat{p}_i=\cos(b_i)\cos(l_i)\,\hat{\bf i}+
 \cos(b_i)\sin(l_i)\,\hat{\bf j}+\sin(b_i)\,\hat{\bf k}\,.
 \ee
 One can find the best-fit dipole direction $(l,\,b)$ and the
 dipole magnitude $A_D$ as well as the other model parameters
 by minimizing the corresponding $\chi^2$.
 Note that in practice $\xi$ can be various observational
 quantities, e.g. the distance modulus $\mu$ of SNIa or
 GRBs~\cite{Mariano:2012wx,Chang:2014nca,Lin:2015rza,Chang:2017bbi,
 Yang:2013gea,Wang:2014vqa}, the centripetal acceleration
 $g_\dag$ in the rotationally supported disk
 galaxies~\cite{Zhou:2017lwy,Chang:2018vxs}, and the varying
 fine structure ``constant'' $\alpha$~\cite{Mariano:2012wx}. We
 refer to e.g.~\cite{Mariano:2012wx,Chang:2014nca,Lin:2015rza,
 Chang:2017bbi,Zhou:2017lwy,Chang:2018vxs,Yang:2013gea,
 Wang:2014vqa} for more details of the DF method.

\begin{center}
 \begin{figure}[tb]
 \centering
 \vspace{-7mm}  
 \includegraphics[width=0.8\textwidth]{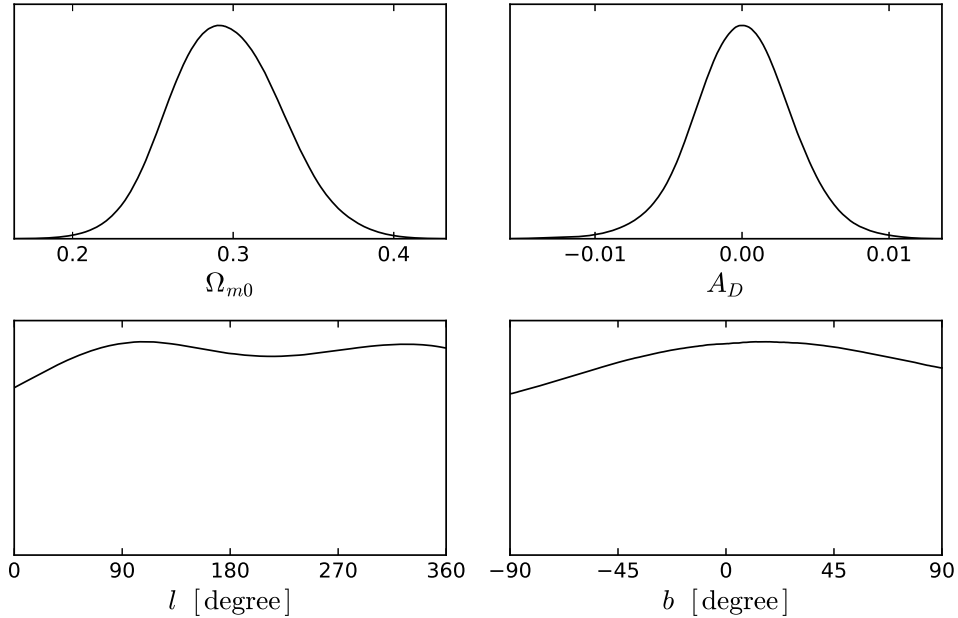}
 \caption{\label{fig3} The marginalized probability distributions of
 $\Omega_{m0}$, the dipole magnitude $A_D$ and the dipole direction
 $(l,\,b)$, obtained by using the DF method to the JLA SNIa dataset
 without any prior on $\Omega_{m0}$ and $A_D$. See the text for
 details.}
 \end{figure}
 \end{center}

In our case of the JLA SNIa dataset, $\xi$ in Eq.~(\ref{eq7})
 is the distance modulus $\mu$ of SNIa. The theoretical
 $\bar{\mu}_{th}$ predicted by the isotropic
 flat $\Lambda$CDM model is given by~\cite{Weinberg:cos,
 Betoule:2014frx,Lin:2015rza,Chang:2017bbi,Zou:2017ksd}
 \be{eq10}
 \bar{\mu}_{th} = 5\log_{10}\frac{d_L}{\rm Mpc} + 25\,,
 \ee
 where the isotropic luminosity distance reads
 \be{eq11}
 d_L(z_{\rm cmb},\,z_{\rm hel})
 =\frac{c\left(1+z_{\rm hel}\right)}{H_0}
 \int_0^{z_{\rm cmb}}\frac{d\tilde{z}}{E(\tilde{z})}\,,
 \ee
 in which $z_{\rm cmb}$ and $z_{\rm hel}$ are the CMB frame redshift
 and heliocentric redshift, respectively; $c$ is the speed of light;
 $H_0$ is the Hubble constant; and
 \be{eq12}
 E(z)=\left[\,\Omega_{m0}(1+z)^3+
 \left(1-\Omega_{m0}\right)\,\right]^{1/2}\,.
 \ee
 We can constrain the dipole direction $(l,\,b)$ and the
 dipole magnitude $A_D$ as well as the flat $\Lambda$CDM model
 parameter $\Omega_{m0}$ by fitting them to the JLA dataset
 consisting of 740 SNIa~\cite{Betoule:2014frx}. Notice that the
 Markov Chain Monte Carlo (MCMC) code CosmoMC~\cite{Lewis:2002ah} is
 used, and the nuisance parameters $H_0$, $\alpha$, $\beta$
 in the distance estimate can be marginalized~\cite{Betoule:2014frx}.
 Following~\cite{Lin:2015rza}, we first require $A_D\geq 0$
 and fix $\Omega_{m0}=0.295$, since the JLA SNIa dataset has
 constrained $\Omega_{m0}=0.295\pm 0.034$ for the isotropic
 flat $\Lambda$CDM model~\cite{Betoule:2014frx}. In Fig.~\ref{fig2},
 we show the marginalized probability distributions of the
 dipole magnitude $A_D$ and the dipole direction $(l,\,b)$. It
 is easy to see that both the distributions of $l$ and $b$ are
 quite flat. This implies that no preferred direction is found.
 In fact, the constraints with $1\sigma$ uncertainties are
 $l={185^\circ}\,_{-185^\circ}^{+175^\circ}\,$,
 $b={5.9^\circ}\,_{-95.9^\circ}^{+84.1^\circ}$ and
 $A_D<3.124\times 10^{-3}$. That is, the $1\sigma$ region of
 $l$ and $b$ is the whole sky ($0^\circ\leq l\leq 360^\circ$,
 $-90^\circ\leq b\leq 90^\circ$), and indeed no preferred direction
 is found. Then, we would like to generalize these results by
 removing the priors $\Omega_{m0}=0.295$ and $A_D\geq 0$
 adopted in~\cite{Lin:2015rza}, namely they are completely free now.
 In this case, we show the marginalized probability distributions of
 $\Omega_{m0}$, the dipole magnitude $A_D$ and the dipole direction
 $(l,\,b)$ in Fig.~\ref{fig3}. Both the distributions of $l$
 and $b$ are still very flat. The constraints with $1\sigma$
 uncertainties are
 $\Omega_{m0}=0.2952\,_{-0.0386}^{+0.0339}\,$,
 $A_D=(0.0\,_{-3.25}^{+3.17})\times 10^{-3}$, and
 $l={179^\circ}\,_{-179^\circ}^{+181^\circ}\,$,
 $b={1.5^\circ}\,_{-91.5^\circ}^{+88.5^\circ}\,$.
 Again, the $1\sigma$ region of $l$ and $b$ is the whole sky
 ($0^\circ \leq l\leq 360^\circ$, $-90^\circ \leq b\leq 90^\circ$),
 and no preferred direction is found by using the DF method.


\vspace{-10mm}  


\section{Comparing two methods by using simulated SNIa
 datasets}\label{sec3}

As is shown in e.g.~\cite{Chang:2014nca,Yang:2013gea,
 Lin:2015rza,Chang:2017bbi} and the previous section, the
 results from the HC method and the DF method are not always
 approximately coincident with each other. If these two methods
 find significantly different preferred directions, which one
 can be trusted? Both or none? If one method finds a preferred
 direction (or more) and the other method finds none, is the
 universe anisotropic or not? In this section, we try to shed
 new light on these questions. Our idea is to test these two
 methods by using several simulated anisotropic SNIa datasets
 with a preset preferred direction or more. We want to see
 which method can find out the preset direction(s), and
 whether the found direction(s) is/are close to the preset
 direction(s). In particular, we try to understand the results
 in Sec.~\ref{sec2}, namely why the DF method fails in the
 JLA SNIa dataset while the HC method works.


\subsection{Methodology to generate the simulated
 SNIa datasets}\label{sec3a}

For simplicity, and without loss of generality, we generate the
 simulated SNIa datasets like the Union2 or Union2.1 SNIa
 datasets, namely the simulated data tables are given directly
 in terms of the distance modulus $\mu$ (with $1\sigma$ uncertainty)
 versus the redshift $z$ of SNIa. Although the JLA/SNLS-like
 simulated SNIa datasets are more complicated mainly due to the
 extra parameters $\alpha$, $\beta$ in the distance estimate,
 the results obtained in this work can be easily extended to
 such kind of simulated datasets.

We take the future SNIa projects in the next decade as a
 reference to generate the simulated SNIa datasets. In this
 regard, the Wide Field Infrared Survey Telescope
 (WFIRST)~\cite{Spergel:2015sza,Spergel:2013uha,WFIRSTwiki,
 Hounsell:2017ejq} to be launched in the mid-2020s might be
 a suitable reference. According to
 e.g.~\cite{Hounsell:2017ejq}, about $3000\sim 8000$ SNIa
 at $z\leq 1.7$ will be available from WFIRST. So, in the
 present work, we will generate $\sim 5000$ simulated SNIa in
 each dataset. Of course, the redshift distribution of SNIa
 tilts to the low-redshift range, and we can use a suitable
 F-distribution~\cite{Fdis} (say, $f(z,\,50,\,0.5)$) to mimic
 the one expected in e.g.~\cite{Hounsell:2017ejq}. According to
 e.g.~\cite{Spergel:2013uha}, the expected aggregate precision
 of these SNIa is $0.20\%$ at $z<1$ and $0.34\%$ at $z>1$.
 Therefore, we assign the simulated $1\sigma$ relative
 uncertainty of the distance modulus $\mu$ to be $0.20\%$ at
 $z<1$ and $0.35\%$ at $z\geq 1$ reasonably.

We generate the distance modulus $\mu$ of the simulated SNIa
 by taking a random number from a Gaussian distribution with
 the mean determined by a flat $\Lambda$CDM model,
 \bea
 &&\disp\mu_{\rm mean}=
 5\log_{10}\frac{d_L}{\rm Mpc}+25\,,\hspace{20mm}
 d_L=\frac{c\left(1+z\right)}{H_0}
 \int_0^z \frac{d\tilde{z}}{E(\tilde{z})}\,,\nonumber\\[1mm]
 && E(z)=\left[\,\Omega_{m0}(1+z)^3+
 \left(1-\Omega_{m0}\right)\,\right]^{1/2}\,,\label{eq13}
 \eea
 where $c$ is the speed of light, and the value of $\Omega_{m0}$ will
 be specified in the particular generating description. The
 Hubble constant $H_0=70\,{\rm km/s/Mpc}$ is adopted as a
 fiducial value, but it does not significantly affect other
 parameters since $H_0$ will be marginalized in fact. The
 standard deviation of this Gaussian distribution is equal to
 the $1\sigma$ uncertainty of $\mu$ mentioned above for the
 particular SNIa, namely $0.20\%$ of $\mu_{\rm mean}$ at $z<1$
 and $0.35\%$ of $\mu_{\rm mean}$ at $z\geq 1$.

Finally, the galactic coordinates $(l,\,b)$ of
 the simulated SNIa will be specified in the particular
 generating description (see below). In fact, the position of
 the simulated SNIa and the value of $\Omega_{m0}$ mentioned
 above will play an important role.


 \begin{center}
 \begin{figure}[tb]
 \centering
 \vspace{-7mm}  
 \includegraphics[width=0.45\textwidth]{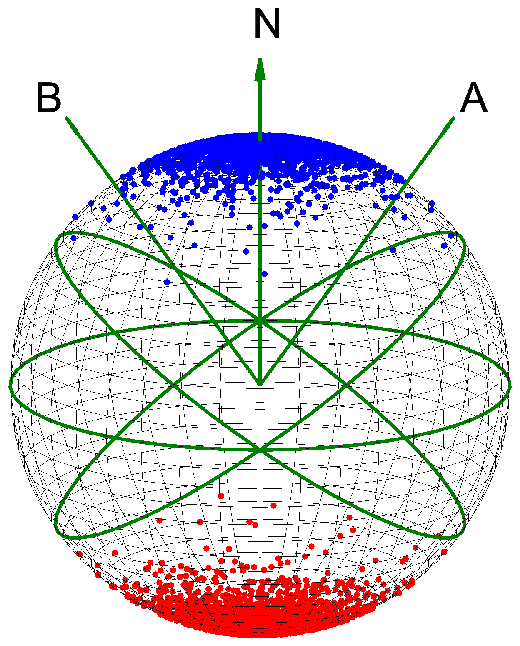}
 \includegraphics[width=0.45\textwidth]{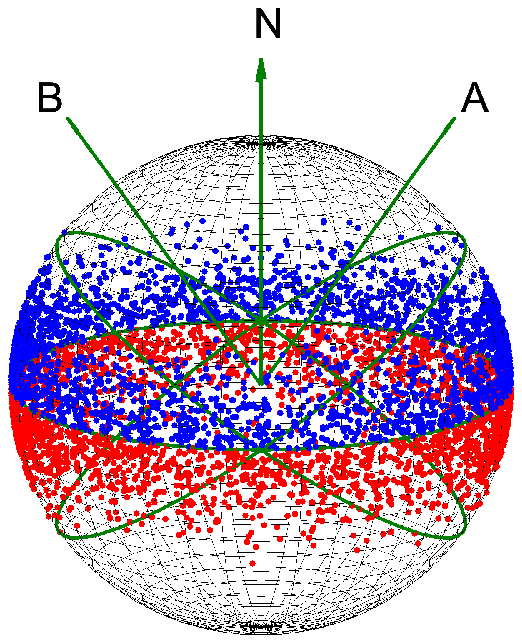}
 \caption{\label{fig4} Demonstration of the spatial distributions of
 the ``Pole-centralized'' (left panel) and
 ``Equator-centralized'' (right panel) simulated SNIa, before
 they are rotated to a preset direction. The blue and red points are
 the SNIa generated with a relatively large and
 small $\Omega_{m0}$, respectively. See the text for details.}
 \end{figure}
 \end{center}


\vspace{-10mm}  


 \begin{center}
 \begin{figure}[tb]
 \centering
 \vspace{-7mm}  
 \includegraphics[width=0.8\textwidth]{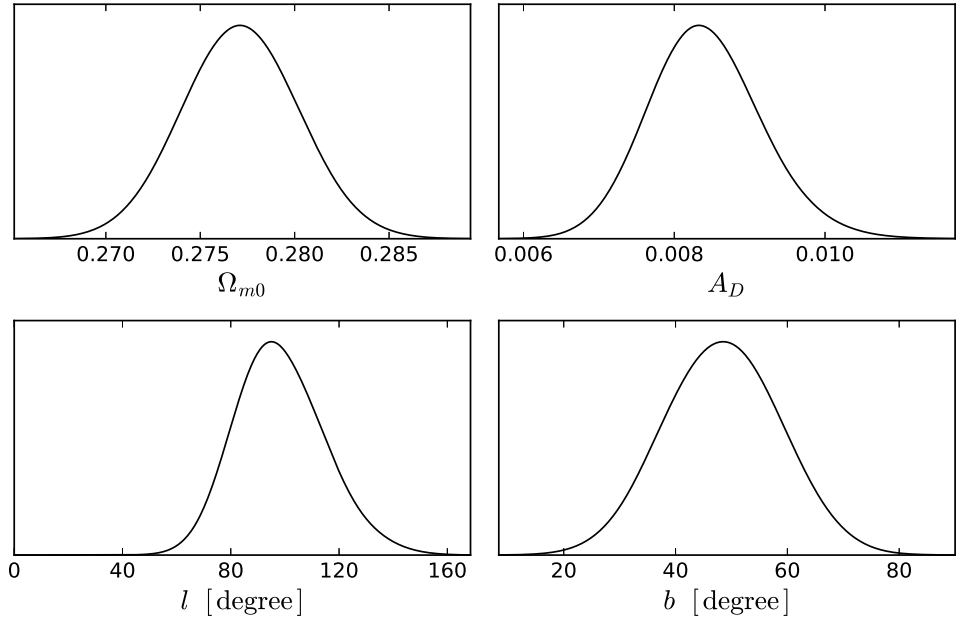}
 \caption{\label{fig5} The marginalized probability distributions of
 $\Omega_{m0}$, the dipole magnitude $A_D$ and the dipole direction
 $(l,\,b)$, obtained by using the DF method to the simulated
 SNIa dataset PC1. See the text for details.}
 \end{figure}
 \end{center}


\vspace{-10mm}  


\subsection{The cases of ``Pole-centralized'' simulated
 SNIa datasets}\label{sec3b}

Let us generate the first simulated SNIa dataset. Here, we
 briefly describe the main steps:

\begin{enumerate}[(P1)]
\setlength{\itemindent}{2.4em}
  \item Construct a Gaussian distribution with the mean at
  the north pole, and a suitable standard deviation
  (say, $30^\circ$). Assign a random number taken from this
  Gaussian distribution to a simulated SNIa as its galactic
  latitude $b$, and assign a random number uniformly taken
  from $[\,0^\circ,\,360^\circ)$ to this simulated SNIa as
  its galactic longitude $l$.
  \item Assign a random redshift from a suitable
  F-distribution (say, $f(z,\,50,\,0.5)$) to
  this simulated SNIa as described in Sec.~\ref{sec3a}.
  \item Generate a distance modulus $\mu$ with $1\sigma$
  uncertainty for this simulated SNIa by using a
  flat $\Lambda$CDM model with a relatively large $\Omega_{m0}$
  (say, $0.45$), as described in Sec.~\ref{sec3a}.
  \item Repeat steps (P1$\sim$3) for 2500 times to generate
  2500 simulated SNIa in the north hemisphere.
  \item Generate 2500 simulated SNIa in the south hemisphere
  with a relatively small $\Omega_{m0}$ (say, $0.15$), similar
  to the previous steps.
  \item By using a suitable coordinate transformation, rotate
  the whole celestial sphere (and all the 5000 simulated SNIa
  adhered to it) to any preset direction (say,
  $(l,\,b)=(120^\circ,\,45^\circ)$).
\end{enumerate}

\noindent When steps (P1$\sim$5) are finished, the sky looks
 like the left panel of Fig.~\ref{fig4}. Clearly, most of the
 simulated SNIa centralize around the north and south poles.
 So, we say such kind of simulated SNIa dataset
 is ``Pole-centralized''. The degree of centralization is
 controlled by the specified standard deviation in step (P1).
 We call the simulated SNIa dataset with the specified
 parameters in the above steps as ``PC1''.

We implement the HC method to the simulated SNIa dataset PC1,
 and repeat 15000 random directions $(l,\,b)$ across the whole
 sky. We find the $1\sigma$ angular region with the maximum AL
 is in the direction
 \be{eq14}
 (l,\,b)\,_{\rm HC}^{\rm PC1}=
 ({138.4632^\circ}\,^{+25.0076^\circ}_{-49.2667^\circ}\,,\,
 {48.5212^\circ}\,^{+19.7711^\circ}_{-25.9446^\circ})\,,
 \ee
 and the corresponding maximum AL (with $1\sigma$
 uncertainty) is
 \be{eq15}
 {\rm AL_{max}^{PC1}}=1.0150\pm 0.0080\,.
 \ee
 Obviously, the direction given in Eq.~(\ref{eq14}) found by the HC
 method is approximately coincident with the preset direction
 $(120^\circ,\,45^\circ)$, but the $1\sigma$ uncertainties
 are fairly large.

Next, we consider the DF method. Noting that $A_D\left(\hat{n}
 \cdot\hat{p}\right)=-A_D\left(-\hat{n}\cdot\hat{p}\right)$ in
 Eq.~(\ref{eq7}), a positive $A_D$ with a direction $\hat{n}$
 is equivalent to a negative $A_D$ with an opposite direction
 $-\hat{n}$. Actually, we have already implemented the DF
 method for many times in various cases, and indeed found two
 peaks in the results, but they are equivalent to each other
 in fact. Therefore, in the rest of this work, without loss of
 generality, we require $A_D\geq 0$
 following e.g.~\cite{Lin:2015rza}.

We implement the DF method with the prior $A_D\geq 0$ to
 the simulated SNIa dataset PC1, and show the marginalized
 probability distributions of $\Omega_{m0}$, the
 dipole magnitude $A_D$ and the dipole direction
 $(l,\,b)$ in Fig.~\ref{fig5}. The constraints
 with $1\sigma$ uncertainties are given by
 \bea
 &&\Omega_{m0}=0.2771^{+0.0031}_{-0.0031}\,,~~~~~~~~~~
 A_D=(8.4164^{+0.6831}_{-0.8011})\times 10^{-3}\,,\label{eq16}\\[2mm]
 &&(l,\,b)\,_{\rm DF}^{\rm PC1}=
 ({98.4333^\circ}\,^{+14.6610^\circ}_{-18.3059^\circ}\,,\,
 {48.4438^\circ}\,^{+10.5712^\circ}_{-10.5792^\circ})\,. \label{eq17}
 \eea
 Although the $1\sigma$ uncertainties are relatively small,
 the direction given in Eq.~(\ref{eq17}) found by the DF method
 deviates from the preset direction $(120^\circ,\,45^\circ)$ beyond
 $1\sigma$ (notice that $l=120^\circ$ is out of the $1\sigma$
 region given in Eq.~(\ref{eq17})). Nonetheless, it is still
 close to the preset direction within the $2\sigma$ region.


 \begin{center}
 \begin{figure}[tb]
 \centering
 \vspace{-7mm}  
 \includegraphics[width=0.8\textwidth]{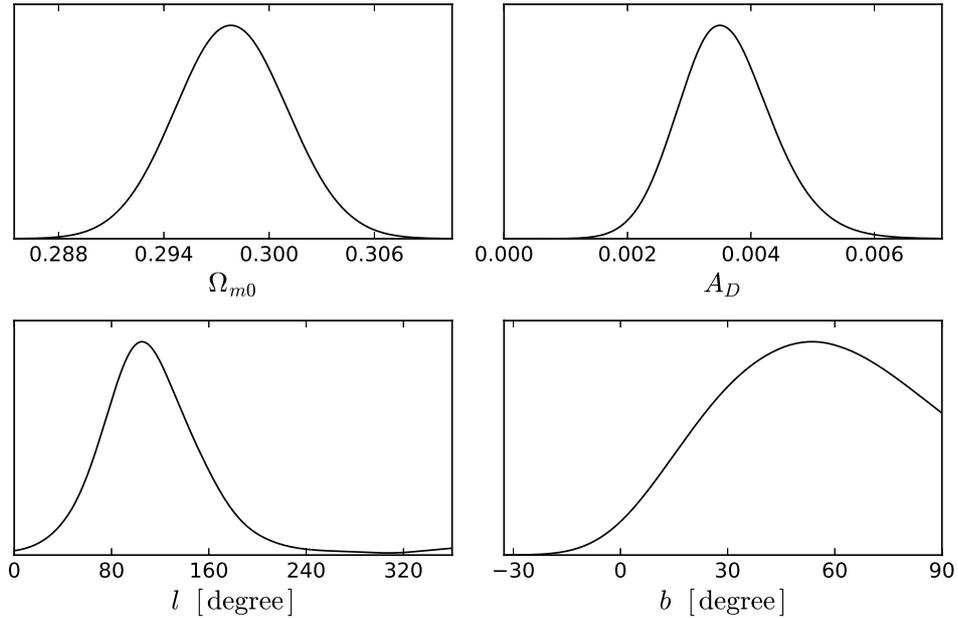}
 \caption{\label{fig6} The same as in Fig.~\ref{fig5}, except
 for the simulated SNIa dataset PC2. See the text for details.}
 \end{figure}
 \end{center}


\vspace{-10mm}  

Noting that in the simulated SNIa dataset PC1, the preset
 ${\rm AL}\sim 2\,(0.45-0.15)/(0.45+0.15)=1$ is fairly high,
 it is natural to see what will happen in the case of lower
 preset AL. So, we generate the second ``Pole-centralized''
 simulated SNIa dataset PC2, by replacing the values of
 $\Omega_{m0}$ in steps (P3) and (P5) with $0.36$ and $0.24$,
 respectively. In this case, the
 preset ${\rm AL}\sim 2\,(0.36-0.24)/(0.36+0.24)=0.4$.

We implement the HC method to the simulated SNIa dataset PC2,
 and repeat 15000 random directions $(l,\,b)$ across the whole
 sky. We find the $1\sigma$ angular region with the maximum AL
 is in the direction
 \be{eq18}
 (l,\,b)\,_{\rm HC}^{\rm PC2}=
 ({116.9953^\circ}\,^{+51.9794^\circ}_{-44.0647^\circ}\,,\,
 {44.7026^\circ}\,^{+34.6490^\circ}_{-23.2385^\circ})\,,
 \ee
 and the corresponding maximum AL (with $1\sigma$
 uncertainty) is
 \be{eq19}
 {\rm AL_{max}^{PC2}}=0.4057\pm 0.0076\,.
 \ee
 Again, the direction given in Eq.~(\ref{eq18}) found by the HC
 method is approximately coincident with the preset direction
 $(120^\circ,\,45^\circ)$, but the $1\sigma$ uncertainties
 are very large.

Then, we implement the DF method with the prior $A_D\geq 0$ to
 the simulated SNIa dataset PC2, and show the marginalized
 probability distributions of $\Omega_{m0}$, the dipole
 magnitude $A_D$ and the dipole direction $(l,\,b)$ in
 Fig.~\ref{fig6}. The constraints with $1\sigma$
 uncertainties are
 \bea
 &&\Omega_{m0}=0.2979^{+0.0031}_{-0.0032}\,,~~~~~~~~~~
 A_D=(3.6069^{+0.6765}_{-0.8190})\times 10^{-3}\,,\label{eq20}\\[2mm]
 &&(l,\,b)\,_{\rm DF}^{\rm PC2}=
 ({116.0269^\circ}\,^{+30.2393^\circ}_{-47.2547^\circ}\,,\,
 {49.3230^\circ}\,^{+30.5256^\circ}_{-20.4601^\circ})\,.\label{eq21}
 \eea
 It is easy to see that the direction given in Eq.~(\ref{eq21})
 found by the DF method is approximately coincident with the
 preset direction $(120^\circ,\,45^\circ)$, but the $1\sigma$
 uncertainties are also very large.

The common feature in the ``Pole-centralized'' simulated SNIa
 datasets is that the uncertainties of the preferred direction
 are fairly large. One can understand this from the left panel
 of Fig.~\ref{fig4}. Since the simulated SNIa centralize around
 two poles, the SNIa located in the ``up hemisphere'' and the
 ``down hemisphere'' are almost the same for e.g.
 the directions A, B and N in the left panel of
 Fig.~\ref{fig4}, although these directions are far from each other.
 Therefore, the values of AL for the directions A, B and N
 are fairly close. As a natural consequence, the preferred
 directions found by both the HC and the DF methods can deviate
 from the preset direction, and the $1\sigma$ angular region
 must be fairly large. Of course, the preset direction is still
 within the $1\sigma$ ($2\sigma$) angular region found by the
 HC (DF) method, while the results of these two methods
 are consistent with each other at the $1\sigma$ level.


 \begin{center}
 \begin{figure}[tb]
 \centering
 \vspace{-8mm}  
 \includegraphics[width=0.8\textwidth]{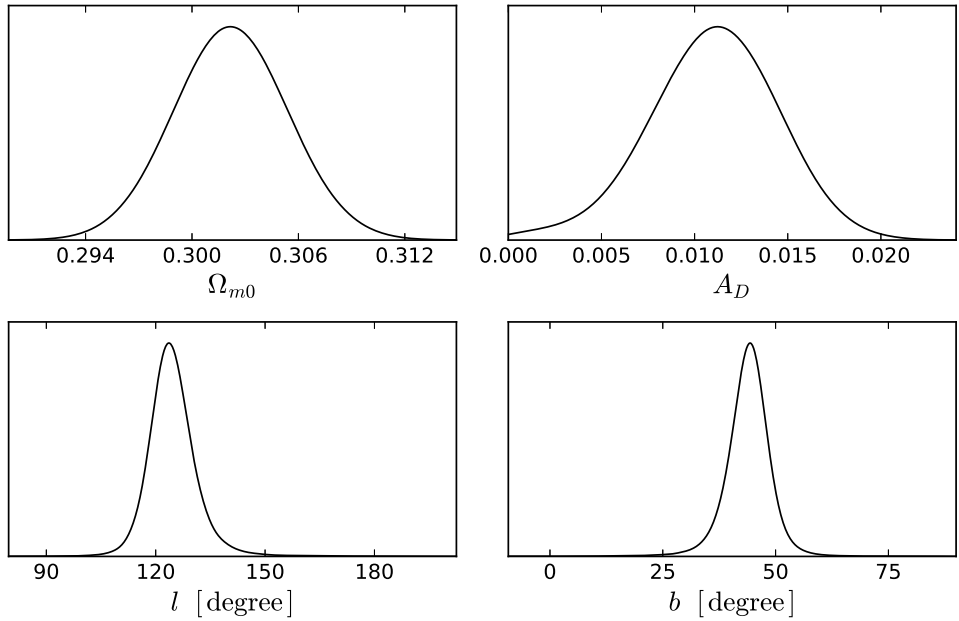}
 \caption{\label{fig7} The same as in Fig.~\ref{fig5}, except
 for the simulated SNIa dataset EC1. See the text for details.}
 \end{figure}
 \end{center}


\vspace{-10mm}  


\subsection{The cases of ``Equator-centralized'' simulated
 SNIa datasets}\label{sec3c}

As is discussed above, the uncertainties of the preferred
 direction in the ``Pole-centralized'' simulated SNIa datasets
 are commonly large. So, we consider another kind of simulated
 SNIa datasets, which are generated in a significantly
 different way. The main steps are

\begin{enumerate}[(E1)]
\setlength{\itemindent}{2.4em}
  \item Construct a Gaussian distribution with the mean at
  the equator (i.e. $b=0$), and a suitable standard deviation
  (say, $10^\circ$). Assign a random number taken from this
  Gaussian distribution to a simulated SNIa as its galactic
  latitude $b$, and assign a random number uniformly taken
  from $[\,0^\circ,\,360^\circ)$ to this simulated SNIa as
  its galactic longitude $l$.
  \item Assign a random redshift from a suitable
  F-distribution (say, $f(z,\,50,\,0.5)$) to
  this simulated SNIa as described in Sec.~\ref{sec3a}.
  \item Generate a distance modulus $\mu$ with $1\sigma$
  uncertainty for this simulated SNIa by using a
  flat $\Lambda$CDM model with a relatively large $\Omega_{m0}$
  (say, $0.36$) if its galactic latitude $b\geq 0$, or with
  a relatively small $\Omega_{m0}$ (say, $0.24$) if its
  galactic latitude $b<0$,  as described in Sec.~\ref{sec3a}.
  \item Repeat steps (E1$\sim$3) for 5000 times
  to generate 5000 simulated SNIa in the whole celestial
  sphere. Notice that the galactic latitudes $b>,\,=,\,<0$
  correspond to the north hemisphere, the equator, the
  south hemisphere, respectively.
  \item By using a suitable coordinate transformation, rotate
  the whole celestial sphere (and all the 5000 simulated SNIa
  adhered to it) to any preset direction (say,
  $(l,\,b)=(120^\circ,\,45^\circ)$).
\end{enumerate}

\newpage  


 \begin{center}
 \begin{figure}[tb]
 \centering
 \vspace{-7mm}  
 \includegraphics[width=0.8\textwidth]{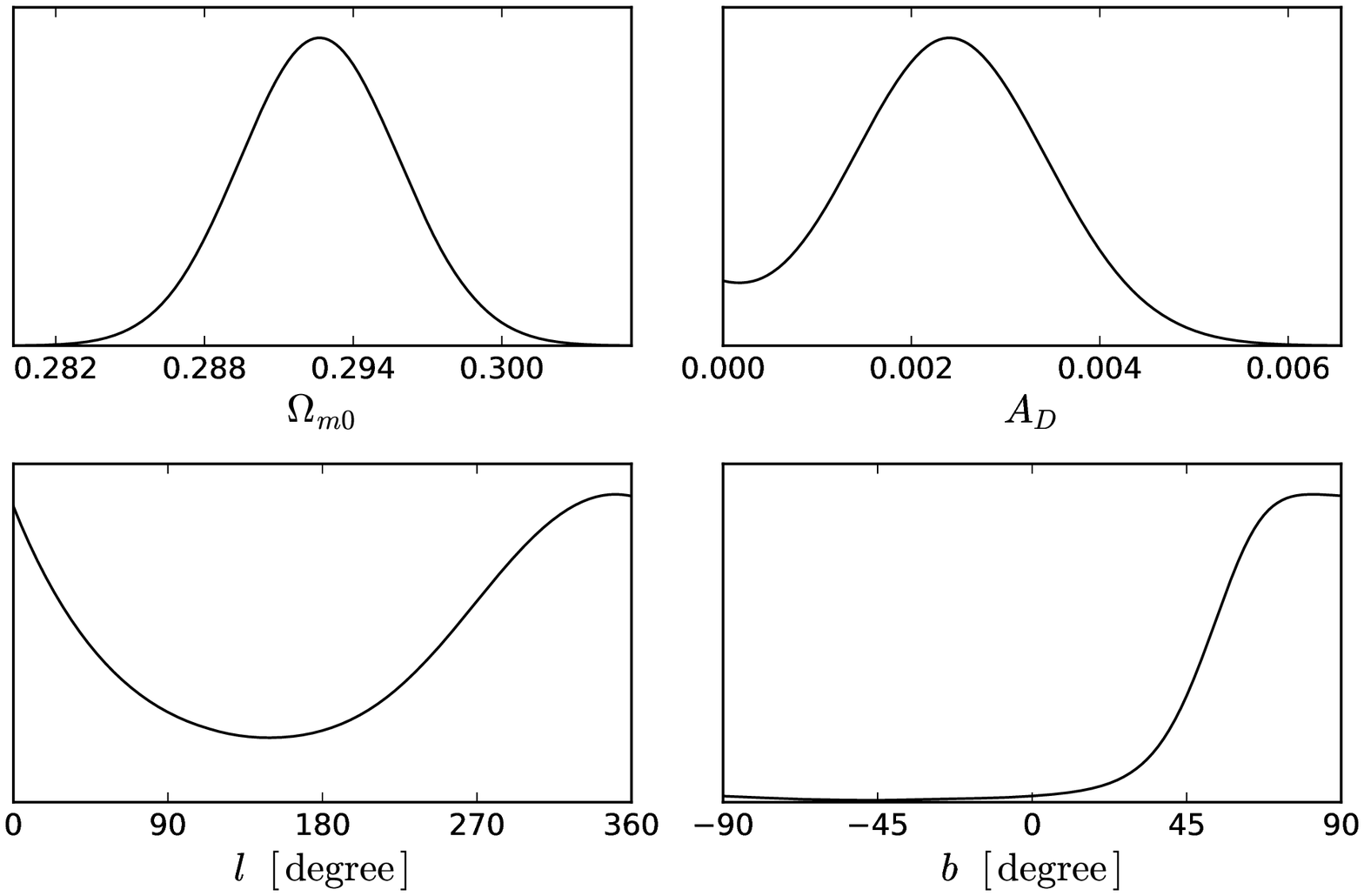}
 \caption{\label{fig8} The same as in Fig.~\ref{fig5}, except
 for the simulated SNIa dataset EC2$_d$. See the text for details.}
 \end{figure}
 \end{center}


\vspace{-8mm}  

\noindent When steps (E1$\sim$4) are finished, the sky looks
 like the right panel of Fig.~\ref{fig4}. Clearly, most of
 the~simulated SNIa centralize around the equator. Thus, we say
 such kind of simulated SNIa dataset
 is ``Equator-centralized''. The degree of centralization is
 controlled by the specified standard deviation in step (E1).
 We call the simulated SNIa dataset with the specified
 parameters in the above steps as ``EC1''.

In contrast to the ``Pole-centralized'' simulated SNIa dataset,
 since the ``Equator-centralized'' simulated SNIa centralize
 around the equator, the SNIa located in the ``up hemisphere''
 and the ``down hemisphere'' for e.g. the directions A and B in the
 right panel of Fig.~\ref{fig4} are significantly different
 from the ones for the direction N. Noting that the blue and
 red points have different $\Omega_{m0}$, it is easy to imagine
 that the directions significantly deviating from the direction
 N will have a much lower AL than the one of the direction N.
 As a natural consequence, the preferred direction found by
 both the HC and the DF methods cannot significantly deviate
 from the preset direction, and the $1\sigma$ angular region
 must be very small.

We implement the HC method to the simulated SNIa dataset EC1,
 and first repeat 15000 random directions $(l,\,b)$ across
 the whole sky. We find that the directions with the largest
 ALs concentrate~around $(121.4857^\circ,\,44.7463^\circ)$, but
 the test random directions within the $1\sigma$ region are
 fairly few. As is discussed above, this is not surprising due
 to the very small $1\sigma$ angular region expected in the cases of
 ``Equator-centralized'' simulated SNIa datasets. Similar to
 the case of JLA SNIa dataset, we densely repeat 5000 random
 directions from a Gaussian distribution with the mean in
 this preliminary direction. Finally, we find the $1\sigma$
 angular region with the maximum AL is in the direction
 \be{eq22}
 (l,\,b)\,_{\rm HC}^{\rm EC1}=
 ({120.2120^\circ}\,^{+0.5732^\circ}_{-1.2585^\circ}\,,\,
 {44.8209^\circ}\,^{+0.9175^\circ}_{-0.2926^\circ})\,,
 \ee
 and the corresponding maximum AL (with $1\sigma$
 uncertainty) is
 \be{eq23}
 {\rm AL_{max}^{EC1}}=0.4138\pm 0.0076\,.
 \ee
 Obviously, the direction given in Eq.~(\ref{eq22}) found by
 the HC method is excellently coincident with the preset
 direction $(120^\circ,\,45^\circ)$, and the $1\sigma$ uncertainties
 are very small, as expected above.

We implement the DF method with the prior $A_D\geq 0$ to
 the simulated SNIa dataset EC1, and show the marginalized
 probability distributions of $\Omega_{m0}$, the
 dipole magnitude $A_D$ and the dipole direction
 $(l,\,b)$ in Fig.~\ref{fig7}. The constraints
 with $1\sigma$ uncertainties are given by
 \bea
 &&\Omega_{m0}=0.3022^{+0.0032}_{-0.0032}\,,~~~~~~~~~~
 A_D=(1.0991^{+0.3668}_{-0.3250})\times 10^{-2}\,,\label{eq24}\\[2mm]
 &&(l,\,b)\,_{\rm DF}^{\rm EC1}=
 ({124.7959^\circ}\,^{+5.0194^\circ}_{-6.7571^\circ}\,,\,
 {43.8235^\circ}\,^{+4.6989^\circ}_{-3.9193^\circ})\,. \label{eq25}
 \eea
 Again, the direction given in Eq.~(\ref{eq25}) found by the DF
 method is approximately coincident with the preset direction
 $(120^\circ,\,45^\circ)$, and the $1\sigma$ uncertainties are
 fairly small, as expected above.

It is easy to see that both the HC and the DF methods work very
 well in the cases of ``Equator-centralized'' simulated SNIa
 datasets. They can find the preset direction correctly, and
 their results are consistent with each other at the $1\sigma$
 level.


 \begin{center}
 \begin{figure}[tb]
 \centering
 \vspace{-9mm}  
 \includegraphics[width=0.8\textwidth]{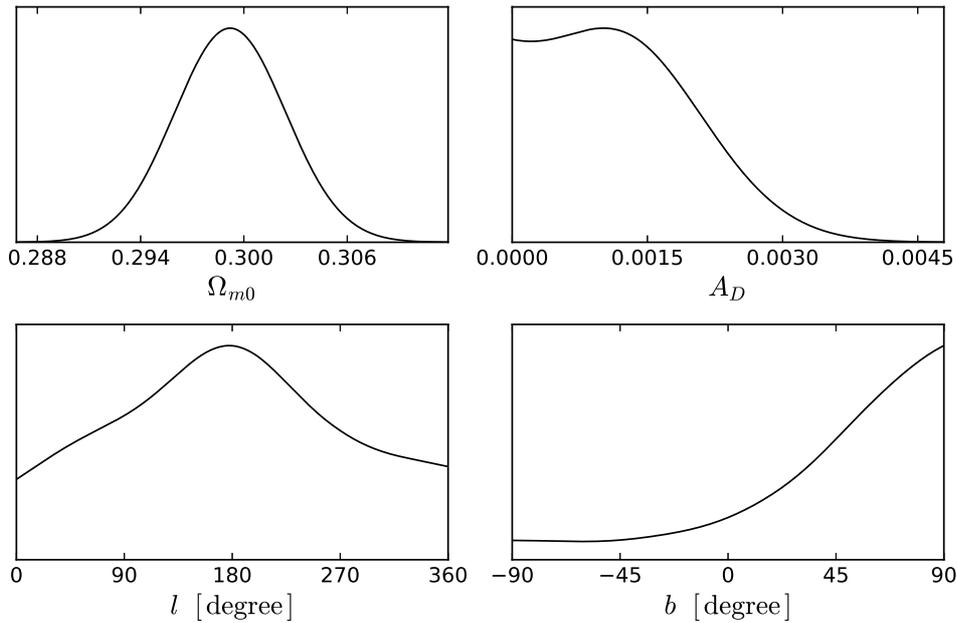}
 \caption{\label{fig9} The same as in Fig.~\ref{fig5}, except
 for the simulated SNIa dataset EC3$_d$. See the text for details.}
 \end{figure}
 \end{center}


\vspace{-10mm}  


\subsection{The cases of simulated SNIa datasets with double
 preset directions}\label{sec3d}

It is suggestive to ponder on the JLA SNIa dataset, where the
 HC method works but the DF method fails. The most noticeable
 feature of the JLA SNIa dataset is that there are two (or even
 more) preferred directions, as is shown in Sec.~\ref{sec2}.
 Therefore, we turn to consider the simulated SNIa datasets
 with double preset directions, which can be easily generated
 by combining two simulated SNIa datasets with different preset
 directions.

As is discussed in Sec.~\ref{sec3c}, the $1\sigma$ uncertainties of
 the preferred direction are fairly small in the case of the
 ``Equator-centralized'' simulated SNIa datasets. So, we choose
 to combine two ``Equator-centralized'' simulated SNIa datasets,
 namely 2500 simulated SNIa with the
 preset direction $(300^\circ,\,45^\circ)$ and another 2500
 simulated SNIa with the preset direction
 $(30^\circ,\,0^\circ)$. Note that the relevant parameters take
 the same values specified in steps (E1$\sim$3). We call the
 resulting simulated SNIa dataset ``EC2$_d$'', which consists
 of 5000 simulated SNIa.

We implement the HC method to the simulated SNIa dataset
 EC2$_d$, and first repeat 15000 random directions $(l,\,b)$
 across the whole sky. We find that the directions with the
 largest ALs concentrate around two directions, i.e.
 $(298.9182^\circ,\,45.2628^\circ)$ and
 $(30.2720^\circ,\,0.2312^\circ)$. Again,
 we densely repeat $5000+5000$ random directions from the
 Gaussian distributions with the means in these two preliminary
 directions, respectively. Finally, we find the $1\sigma$
 angular region with the maximum AL is in the direction
 \be{eq26}
 (l,\,b)\,_{\rm HC,\,max}^{{\rm EC2}_d} =
 ({29.8852^\circ}\,^{+0.4980^\circ}_{-0.3569^\circ}\,,\,
 {-0.1492^\circ}\,^{+0.3091^\circ}_{-0.2171^\circ})\,,
 \ee
 and the corresponding maximum AL (with $1\sigma$
 uncertainty) is
 \be{eq27}
 {\rm AL}_{\rm max}^{{\rm EC2}_d}=0.2206\pm 0.0078\,.
 \ee
 In addition, we also find a sub-maximum AL in the direction
 (with $1\sigma$ uncertainty)
 \be{eq28}
 (l,\,b)\,_{\rm HC,\,sub}^{{\rm EC2}_d} =
 ({299.9246^\circ}\,^{+1.0814^\circ}_{-1.4783^\circ}\,,\,
 {44.5268^\circ}\,^{+0.7310^\circ}_{-1.3538^\circ})\,,
 \ee
 and the corresponding sub-maximum AL (with $1\sigma$ uncertainty) is
 \be{eq29}
 {\rm AL}_{\rm sub}^{{\rm EC2}_d}=0.1538\pm 0.0077\,.
 \ee
 Clearly, these two preferred directions given in
 Eqs.~(\ref{eq26}) and (\ref{eq28}) found by the HC method are
 excellently coincident with the two preset directions
 $(30^\circ,\,0^\circ)$ and $(300^\circ,\,45^\circ)$, while the
 $1\sigma$ uncertainties are very small, as expected above.
 The HC method works very well.

We implement the DF method with the prior $A_D\geq 0$ to
 the simulated SNIa dataset EC2$_d$, and show the marginalized
 probability distributions of $\Omega_{m0}$, the
 dipole magnitude $A_D$ and the dipole direction
 $(l,\,b)$ in Fig.~\ref{fig8}. The constraints
 with $1\sigma$ uncertainties are given by
 \bea
 &&\Omega_{m0}=0.2927^{+0.0032}_{-0.0032}\,,~~~~~~~~~~
 A_D=(2.3638^{+1.0354}_{-1.0541})\times 10^{-3}\,,\label{eq30}\\[2mm]
 &&(l,\,b)\,_{\rm DF}^{{\rm EC2}_d}=
 ({183.4706^\circ}\,^{+176.5294^\circ}_{-183.4706^\circ}\,,\,
 {64.1008^\circ}\,^{+25.8992^\circ}_{-4.1372^\circ})\,. \label{eq31}
 \eea
 Obviously, the DF method cannot correctly find out any one
 of the two preset directions $(300^\circ,\,45^\circ)$ and
 $(30^\circ,\,0^\circ)$, while the $1\sigma$ uncertainties
 are very large. In fact, the $1\sigma$ regions of $l$ and $b$
 are $0^\circ\leq l\leq 360^\circ$ and
 $59.9636^\circ\leq b\leq 90^\circ$, respectively. That is, it
 ``finds'' a very wide $1\sigma$ angular region, which is not
 true in fact. The DF method fails in this case.

Further, we consider a fairly different case, in which the
 simulated SNIa are more centralized around the equator. We
 generate another simulated SNIa dataset EC3$_d$, which is the
 same as EC2$_d$ but the specified standard deviation in step
 (E1) is replaced with $2^\circ$.

We implement the HC method to the simulated SNIa dataset
 EC3$_d$, and first repeat 15000 random directions $(l,\,b)$
 across the whole sky. We find that the directions with the
 largest ALs concentrate around two directions, i.e.
 $(299.5397^\circ,\,45.6902^\circ)$ and
 $(30.7204^\circ,\,-0.5147^\circ)$. Again,
 we densely repeat $5000+5000$ random directions from the
 Gaussian distributions with the means in these two preliminary
 directions, respectively. Finally, we find the $1\sigma$
 angular region with the maximum AL is in the direction
 \be{eq32}
 (l,\,b)\,_{\rm HC,\,max}^{{\rm EC3}_d} =
 ({30.1148^\circ}\,^{+0.1339^\circ}_{-0.3388^\circ}\,,\,
 {-0.1085^\circ}\,^{+0.1916^\circ}_{-0.2465^\circ})\,,
 \ee
 and the corresponding maximum AL (with $1\sigma$
 uncertainty) is
 \be{eq33}
 {\rm AL}_{\rm max}^{{\rm EC3}_d}=0.2058\pm 0.0076\,.
 \ee
 In addition, we also find a sub-maximum AL in the direction
 (with $1\sigma$ uncertainty)
 \be{eq34}
 (l,\,b)\,_{\rm HC,\,sub}^{{\rm EC3}_d} =
 ({299.9898^\circ}\,^{+0.2998^\circ}_{-0.1614^\circ}\,,\,
 {44.9113^\circ}\,^{+0.3212^\circ}_{-0.0486^\circ})\,,
 \ee
 and the corresponding sub-maximum AL (with $1\sigma$ uncertainty) is
 \be{eq35}
 {\rm AL}_{\rm sub}^{{\rm EC3}_d}=0.1738\pm 0.0076\,.
 \ee
 In this case, these two preferred directions given in
 Eqs.~(\ref{eq32}) and (\ref{eq34}) found by the HC method are
 still excellently coincident with the two preset directions
 $(30^\circ,\,0^\circ)$ and $(300^\circ,\,45^\circ)$, while the
 $1\sigma$ uncertainties are very small. The HC method still
 works very well.

We implement the DF method with the prior $A_D\geq 0$ to
 the simulated SNIa dataset EC3$_d$, and show the marginalized
 probability distributions of $\Omega_{m0}$, the
 dipole magnitude $A_D$ and the dipole direction
 $(l,\,b)$ in Fig.~\ref{fig9}. The constraints
 with $1\sigma$ uncertainties are given by
 \bea
 &&\Omega_{m0}=0.2992^{+0.0032}_{-0.0032}\,,~~~~~~~~~~
 A_D=(1.2781^{+0.3608}_{-1.2707})\times 10^{-3}\,,\label{eq36}\\[2mm]
 &&(l,\,b)\,_{\rm DF}^{{\rm EC3}_d}=
 ({181.1659^\circ}\,^{+100.9199^\circ}_{-100.0070^\circ}\,,\,
 {42.6214^\circ}\,^{+47.3786^\circ}_{-7.9784^\circ})\,. \label{eq37}
 \eea
 Again, the DF method cannot correctly find out any one
 of the two preset directions $(300^\circ,\,45^\circ)$ and
 $(30^\circ,\,0^\circ)$, while the $1\sigma$ uncertainties
 are very large. In fact, the $1\sigma$ regions of $l$ and $b$
 are $81.1589^\circ\leq l\leq 282.0858^\circ$ and
 $34.6430^\circ\leq b\leq 90^\circ$, respectively. That is,
 it ``finds'' a very wide but wrong $1\sigma$ angular region.
 The DF method fails once more.

Clearly, the DF method cannot find any preset directions in
 the above two cases. Thus, we conclude that the DF method
 cannot properly work in the SNIa datasets with two (or even
 more) preferred directions, while the HC method still works
 well. In particular, this might help us to understand the
 results in the JLA SNIa dataset (see Sec.~\ref{sec2}).
 Briefly, as is found in Sec.~\ref{sec2}, there exist at least
 two preferred directions $(23.49^\circ,\,2.25^\circ)$ and
 $(299.47^\circ,\,28.39^\circ)$ in the JLA SNIa dataset, and
 hence it is not surprising that the DF method fails in this
 case. The JLA SNIa dataset is indeed a realistic example
 to show the shortcoming of the DF method.


 \begin{table}[htb]
 \renewcommand{\arraystretch}{1.45}
 \begin{center}
 \vspace{4mm} 
 \begin{tabular}{lcl} \hline\hline
 \ Dataset  & \quad Preferred direction $(l,\,b)$ \quad\quad\quad &  Ref.  \\ \hline
 \ Union2 SNIa (HC)  &  $(309^\circ,\,18^\circ)$ & \cite{Antoniou:2010gw} \ \\
 \ Union2 SNIa (DF)  &  $(309^\circ,\,-15^\circ)$ & \cite{Mariano:2012wx} \ \\
 \ Union2.1 SNIa (DF)  &  $(307^\circ,\,-14^\circ)$ & \cite{Yang:2013gea} \ \\
 \ CMB Dipole  &  $(264^\circ,\,48^\circ)$ & \cite{Lineweaver:1996xa} \ \\
 \ Velocity Flows & $(282^\circ,\,6^\circ)$ & \cite{Feldman:2009es} \ \\
 \ Quasar Alignment & $(267^\circ,\,69^\circ)$ & \cite{Hutsemekers} \ \\
 \ GRBs\,+\,Union2.1 SNIa (DF) & $(309^\circ,\,-8.6^\circ)$ & \cite{Wang:2014vqa} \ \\
 \ $\Delta\alpha/\alpha$ & $(330^\circ,\,-13^\circ)$ & \cite{King:2012id,Webb:2010hc} \ \\
 \ CMB Quadrupole & $(240^\circ,\,63^\circ)$ & \cite{Frommert:2009qw,Bielewicz:2004en} \ \\
 \ CMB Octopole & $(308^\circ,\,63^\circ)$ & \cite{Bielewicz:2004en} \ \\
 \ SPARC Galaxies (HC max.) & $(175.5^\circ,\,-6.5^\circ)$ & \cite{Zhou:2017lwy} \ \\
 \ SPARC Galaxies (HC sub-max.) & $(114.5^\circ,\,2.5^\circ)$ & \cite{Zhou:2017lwy} \ \\
 \ SPARC Galaxies (DF) & $(171^\circ,\,-15^\circ)$ & \cite{Chang:2018vxs} \ \\
 \ JLA SNIa (HC max.) & $(23.49^\circ,\,2.25^\circ)$ & This work \ \\
 \ JLA SNIa (HC sub-max.) & $(299.47^\circ,\,28.39^\circ)$ & This work \ \\
 \hline\hline
 \end{tabular}
 \end{center}
 \caption{\label{tab1} Preferred directions $(l,\,b)$ found in
 various observational datasets.}
 \end{table}


\vspace{-5mm}  


\section{Concluding remarks}\label{sec4}

The cosmological principle is one of the cornerstones in modern
 cosmology~\cite{Weinberg:cos,Kolb90}. It assumes that the
 universe is homogeneous and isotropic on cosmic scales. Both
 the homogeneity and the isotropy of the universe should be
 tested carefully. In the present work, we are interested in
 probing the possible preferred direction in the distribution
 of SNIa. To our best knowledge, two main methods have been
 used in almost all of the relevant works in the literature,
 namely the HC method and the DF method. However, the results
 from these two methods are not always approximately coincident
 with each other. In this work, we test the cosmic anisotropy
 by using these two methods with the JLA and simulated SNIa
 datasets. In many cases, both methods work well, and their
 results are consistent with each other. However, in the cases
 with two (or even more) preferred directions, the DF method
 fails while the HC method still works well. This might
 shed new light on our understanding of these two methods.

\begin{center}
 \begin{figure}[tb]
 \centering
 \vspace{-2mm}  
 \includegraphics[width=0.85\textwidth]{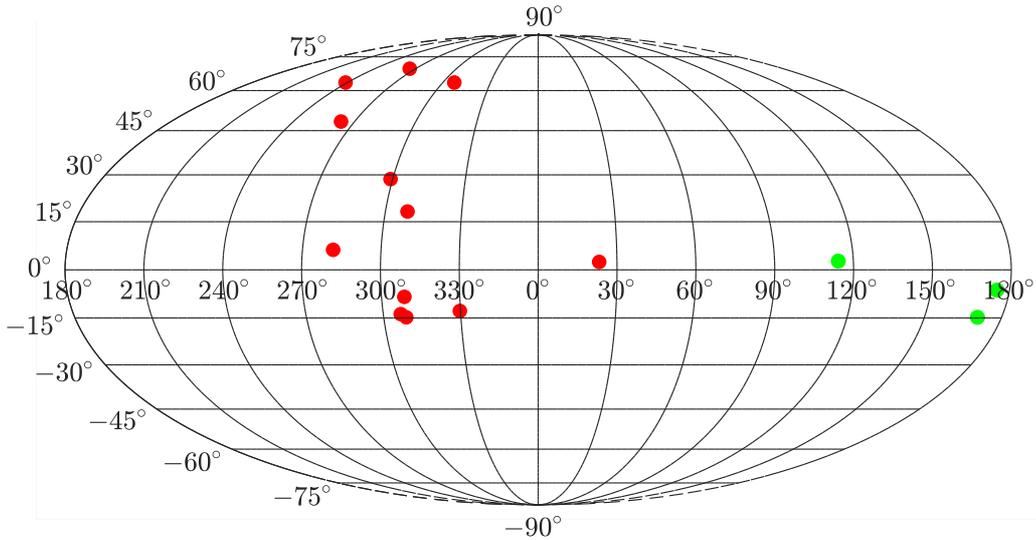}
 \caption{\label{fig10} Preferred directions $(l,\,b)$ found in
 various observational datasets (see Table~\ref{tab1}). Note
 that the three preferred directions in the SPARC Galaxies are
 labeled by the green points, while the others are labeled
 by the red points. See the text and Table~\ref{tab1} for details.}
 \end{figure}
 \end{center}

In Table~\ref{tab1}, we summarize the preferred directions
 $(l,\,b)$ found in various observational datasets. We also
 plot them in Fig.~\ref{fig10}. Most of them (including the
 two preferred directions of the JLA SNIa dataset found in
 this work) are located in a relatively small part (about a
 quarter) of the north galactic hemisphere, as is shown by
 the red points in Fig.~\ref{fig10}. In some sense, they are
 in agreement with each other. However, the three preferred
 directions found in the SPARC Galaxies are significantly
 different from the others, as is shown by the green points in
 Fig.~\ref{fig10}. Note that these preferred directions in the
 SPARC Galaxies are found by using the centripetal acceleration
 $g_\dag$~\cite{Zhou:2017lwy,Chang:2018vxs}. This is different
 from the others, and might be responsible for the difference.
 Nevertheless, we stress that the two preferred directions of
 the JLA SNIa dataset found in this work are
 clearly in agreement with the other ten preferred directions.


\vspace{-8mm}  

Several remarks are in order.
 In this work, we only consider the spatially flat $\Lambda$CDM
 model. In fact, one can generalize our discussions to other
 cosmological models, such as $w$CDM, CPL models. Of course,
 one can also consider model-independent parameterizations.
 It is reasonable to expect that our results do not change
 significantly in these generalized cases.

In the HC method, we have used $\Omega_{m0}$ (equivalent to the
 accelerating expansion rate, namely the deceleration parameter
 $q_0$, in the spatially flat $\Lambda$CDM model) to define AL,
 as in Eq.~(\ref{eq1}). In fact, one can instead define AL by using
 other quantities characterizing the cosmic expansion, e.g. the
 deceleration parameter $q_0$ directly~\cite{Cai:2011xs} and
 the Hubble rate $H_0$~\cite{Chang:2014nca}.

Here, we have only considered the SNIa datasets. In fact, one
 can extend our work to the data of other observations, such as
 GRBs~\cite{Meszaros:2009ux,Wang:2014vqa,Chang:2014jza,Liu:2014vda},
 rotationally supported galaxies~\cite{Zhou:2017lwy,Chang:2018vxs},
 quasars and radio galaxies~\cite{Singal:2013aga}, quasar
 optical polarization data~\cite{Hutsemekers,Pelgrims:2016mhx},
 and the varying fine structure ``constant''
 $\alpha$~\cite{Mariano:2012wx}.

In this work, we have only considered two kinds of simulated
 SNIa datasets, which are ``Pole-centralized''
 and ``Equator-centralized'', respectively. In fact, one can
 further consider other kinds of simulated SNIa datasets. The
 distribution of simulated SNIa can be more general. On the
 other hand, one can further consider the simulated SNIa
 datasets with three or four preset directions to test both
 the HC method and the DF method.

Here are further discussions on the failure of the DF method
 in the cases with two (or even more) preferred directions.
 Since the DF method only models a single dipole, this failure
 is not surprising in fact. It is of interest to test the DF
 method and the HC method by using the simulated SNIa datasets
 with multiple dipoles of different amplitudes (we thank the
 anonymous referee~A for pointing out this issue). However,
 we admit that it is fairly difficult to generate such kind
 of simulated SNIa datasets, and some smart ideas are needed
 to this end. We leave it to future work. On the other hand,
 the DF method might be improved by adding a quadrupole term
 in Eq.~(\ref{eq7}), or by simply generalizing the angular
 dependent function, e.g. replacing $\hat{n}\cdot\hat{p}$ with
 a function of $\hat{n}\cdot\hat{p}$ (we thank the anonymous
 referee~B for pointing out this issue). In addition, although
 the monopole term does not encode the information of anisotropy and
 is indeed negligible in most of the relevant works, it is still of
 interest to identify the corresponding effect in the context
 of the DF method (we thank again the anonymous referee~B for
 pointing out this issue). Since both improvements will
 remarkably extend the length of this paper,
 we hope to consider these issues in future work.

Although we have shown that the HC method works well while the
 DF method might fail in some complicated cases, it does not
 mean that one should not continue to use the DF method in the
 relevant works. Actually, the DF method works well in most
 cases and the corresponding results are approximately
 coincident with the ones of the HC method. Most importantly,
 the DF method is more efficient than the~HC~method, namely
 it consumes less computational power and time. In the HC
 method, in order to find the preferred direction precisely,
 one needs to significantly increase the number of the random
 directions in searching the direction with the maximum AL. For
 example, it took more than 1 week to calculate the ALs for
 $\sim 40000$ random directions in the JLA SNIa dataset (see
 Sec.~\ref{sec2}) by using our computer. However, employing
 the MCMC code CosmoMC~\cite{Lewis:2002ah} instead, the DF
 method only took $\sim 10$ hours to obtain the satisfactory
 result by using the same computer. The algorithm of the DF
 method makes it more efficient than the HC method, and hence
 the DF method is still a valuable tool in the relevant works.

Since they have been used extensively in the literature, we
 consider that both the HC method and the DF method need to
 be improved. Further corrections or even completely new
 methods are desirable. New ideas are welcome.


\section*{ACKNOWLEDGEMENTS}

We thank the anonymous referees for quite useful comments and
 suggestions, which helped us to improve this work. We are grateful
 to Xiao-Bo~Zou, Shou-Long~Li, Zhao-Yu~Yin, Dong-Ze~Xue,
 Da-Chun~Qiang, and Zhong-Xi~Yu for kind help and discussions. This
 work was supported in part by NSFC under Grants No.~11575022
 and No.~11175016.

\renewcommand{\baselinestretch}{1.01}


\end{document}